\documentclass[letter,twocolumn]{jpsj3}

\title{Calibration of the Particle Density in Cellular-Automaton Models for Traffic Flow}

\author{Masahiro Kanai\thanks{E-mail address: kanai@ms.u-tokyo.ac.jp} %\\
}
\inst{Graduate School of Mathematical Sciences, The University of Tokyo, Komaba 3-8-1, Meguro-ku, Tokyo%\\
}
\abst{
We introduce density dependence of the cell size in cellular-automaton models for traffic flow, which allows a more precise correspondence between real-world phenomena and what observed in simulation.
Also, we give an explicit calibration of the particle density particularly for the asymmetric simple exclusion process with some update rules.
We thus find that the present method is valid in that it reproduces a realistic flow-density diagram.
}

\kword{traffic flow, cellular automaton, stochastic model, asymmetric simple exclusion process, exact solution}

\begin{document}
\maketitle
%%%%%%%%%%%%%%%%%%%%%%%%%%%%%%%%%%%%%%%%%%%%%%%%%%%
In the last several decades, simulation research has firmly established its place in physics.
Especially, {\it cellular automaton} (CA) modeling plays a central role in hydrodynamics, statistical mechanics and general phenomenological studies including complex systems \cite{Wolfram,DeMasi,Lebowitz}.
CA models are, naturally, suitable for simulation on a computer, but there are also actual discrete systems like CA, e.g., molecular motors on a microtubule \cite{Howard}.
In such systems, particles have a finite volume, and in principle, both collision and overtaking are forbidden.
As a result, we often observe some particles stemming the flow by themselves.
The above situation can be precisely mimicked with the so-called {\it exclusion rule} in CA models.

Introduction of CA models produced a breakthrough in research on traffic flow \cite{Chowdhury,Helbing,Nagel,Fukui,Nagel2,Knospe,Kanai1,Kanai5}, and now CA models are widely applied in real-world technologies, such as in real-time traffic forecast.
One of the most basic models for traffic flow is the asymmetric simple exclusion process \cite{MacDonald,Spitzer} (ASEP), a CA model incorporating randomness into an elementary CA rule 184 \cite{Wolfram}; particles in the ASEP hop to the next site with a given probability if it is not occupied.
We should remark here that there are some possible choices to update cells \cite{Rajewsky,Kanai2}, e.g., random sequential updating and shuffled dynamics \cite{Wolki}, whereas all sites are synchronously updated in a CA \cite{Wolfram}.
However, we do not adhere to only the parallel update case in the following discussion.

The validation of traffic-flow models is carried out by comparing the simulation result to observational data \cite{Nagel,Nagel2,Knospe,Bando,Kerner,Tadaki,Sugiyama}.
Normal traffic data contains the density, velocity and flow, and one usually plots the velocity against the flow, the velocity against the density and the flow against the density.
In particular, the significant properties of a traffic flow are summed up in the flow-density diagram \cite{Lighthill}.
Some simple models such as the ASEP are known to be exactly solvable \cite{Rajewsky,Kanai2,Kanai3,Schadschneider}, which means that one can obtain an exact flow-density diagram.
Note that the flow-density diagram for the ASEP has the so-called hole-particle symmetry: i.e., inverting all the cells in concert, one sees that the diagram is symmetric in the density.
In what follows, we discuss a calibration of CA models for traffic flow focusing on these exactly solvable ones.

In this work, we simply consider $\nu$ vehicles of the same size $c$ on a circuit of length $l$.
The {\it real-world} density of vehicles, which is conserved in simulations, is accordingly defined as
\begin{equation}
\rho_{\mbox{\tiny RW}}=\frac{\nu c}l\qquad (0\leq\rho_{\mbox{\tiny RW}}\leq1).
\end{equation}
Meanwhile, in the corresponding CA model the particle density is defined as
\begin{equation}
\rho_{\mbox{\tiny CA}}=\frac{N}{L}\qquad (0\leq\rho_{\mbox{\tiny CA}}\leq1),
\end{equation}
where $N$ is the number of occupied cells, and $L$ is the total number of cells.
Let $d$ be the cell size. Since $\nu=N$ and $l=Ld$, we then have
\begin{equation}
\frac{\rho_{\mbox{\tiny CA}}}{\rho_{\mbox{\tiny RW}}}=\frac{d}{c}.\label{rhos}
\end{equation}
In this letter, we shed light on the following condition with respect to vehicle size $c$ and cell size $d$:
\begin{equation}
c\leq d\leq 2c.\label{condition}
\end{equation}
This is justified because if $d<c$ then vehicles could not be contained in each single cell, or if $2c<d$ then one should reduce the cell size $d$ to half.
The CA models for traffic flow admit of optimizing the cell size according to the particle density.
We therefore assume that the cell size $d$ is determined depending on the density $\rho_{\mbox{\tiny RW}}$ as $d=cf(\rho_{\mbox{\tiny RW}})$, where $f(\rho)$ is a function such that $1\leq f(\rho)\leq 2$ due to eq. (\ref{condition}).
Accordingly, eq. (\ref{rhos}) yields
\begin{equation}
\rho_{\mbox{\tiny CA}}=\rho_{\mbox{\tiny RW}}f(\rho_{\mbox{\tiny RW}}).
\end{equation}
Note that since $0\leq \rho_{\mbox{\tiny CA}}\leq 1$ by definition, we have to refine the condition on $f(\rho)$ to
\begin{equation}
1\leq f(\rho)\leq\min\{2,1/\rho\}.
\end{equation}
One may use a larger cell size while the density is low, but a smaller one while it is high.
Thus, we define $f(\rho)$ as a non-increasing function in $\rho$.
For example, a simple choice, $f(\rho)=2-\rho$, leads to
\begin{gather}
\rho_{\mbox{\tiny CA}}=\rho_{\mbox{\tiny RW}}(2-\rho_{\mbox{\tiny RW}}).\label{calib}
\end{gather}

As mentioned above, the ASEP is exactly solvable.
The flow $Q_{\mbox{\tiny random}}$ is described as a function of the density $\rho_{\mbox{\tiny CA}}$:
\begin{equation}
Q_{\mbox{\tiny random}}(\rho_{\mbox{\tiny CA}})=p\rho_{\mbox{\tiny CA}}(1-\rho_{\mbox{\tiny CA}}),\label{ASEPr}
\end{equation}
where $p$ is the hop probability.
Note that eq. (\ref{ASEPr}) is for the random sequential updating.
Then, using eq. (\ref{calib}), the flow-density diagram, eq. (\ref{ASEPr}), is calibrated as
\begin{gather}
Q_{\mbox{\tiny random}}(\rho_{\mbox{\tiny RW}})=p\rho_{\mbox{\tiny RW}}(2-\rho_{\mbox{\tiny RW}})(1-\rho_{\mbox{\tiny RW}})^2.\label{ASEPrc}
\end{gather}
(Note that we often denote different functions by the same symbol for the sake of simplicity.)
In Fig. 1(a), we show the graphs of eqs. (\ref{ASEPr}) and (\ref{ASEPrc}), finding that the calibration (\ref{calib}) helps to reproduce a more realistic flow-density diagram; in particular, the maximum-flow density moves from $1/2$ to $1-1/\sqrt{2}~(\simeq0.29)$.

Also, we consider the average velocity of vehicles as a function of the density.
Following the assumption~\cite{Lighthill}, $Q=\rho v$, with respect to flow $Q$, density $\rho$, and velocity $v$, we have
\begin{equation}
\begin{aligned}
v_{\mbox{\tiny random}}(\rho_{\mbox{\tiny CA}})=&p(1-\rho_{\mbox{\tiny CA}}),\\
v_{\mbox{\tiny random}}(\rho_{\mbox{\tiny RW}})=&p(2-\rho_{\mbox{\tiny RW}})(1-\rho_{\mbox{\tiny RW}})^2.
\label{ASEPvr}
\end{aligned}
\end{equation}
Figure 1(b) shows the graphs of eq. (\ref{ASEPvr}).
We see density dependence of the mean hop probability in the ASEP.
Note that after the calibration the average velocity is no longer equivalent to the mean hop probability and can take values larger than 1.
%, which reveals how the proposed calibration works.
% the dynamics mimicked by the simple particle-hopping model with a constant hop probability and moreover how the proposed calibration works.
%Once the cell size is fixed, dynamics of the {\it real-world} vehicles is transformed into a coarse-grained motion in the CA model: Particles hop at most one site with a constant probability and the exclusive interaction.
%The calibration then helps one restore the original dynamics.
%%%%%%%%%
\begin{figure}[]
%\begin{flushright}
\hspace{15pt}\includegraphics[scale=1.]{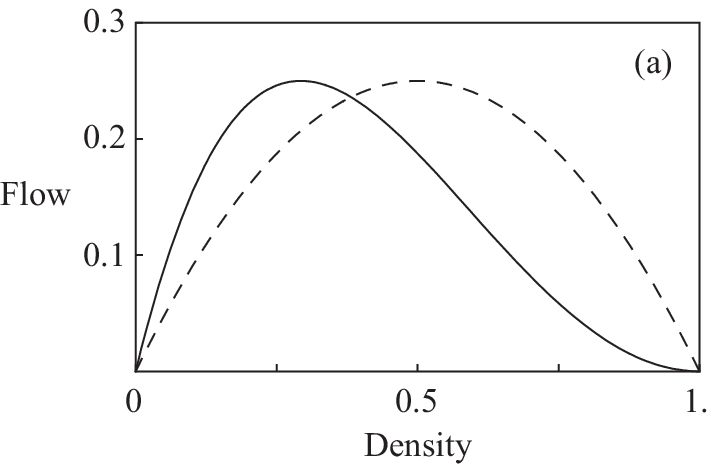}\\
\hspace{6pt}\includegraphics[scale=1.]{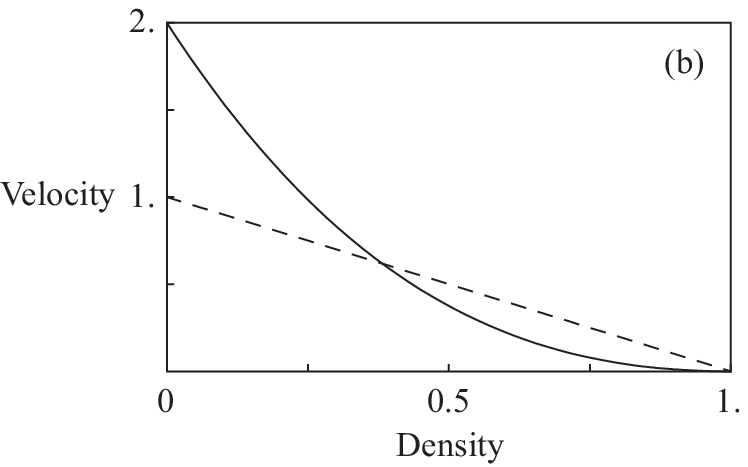}
\caption{
The ASEP with random sequential updating and hop probability $p=1$: (a)The flow-density diagram and (b)The velocity-density diagram; (solid line)calibrated and (dashed line)non-calibrated.
}
\label{fig1}
%\end{flushright}
\end{figure}
%%%%%%%%%
%%%%%%%%%
\begin{figure}[]
%\begin{center}
\hspace{15pt}\includegraphics[scale=1.]{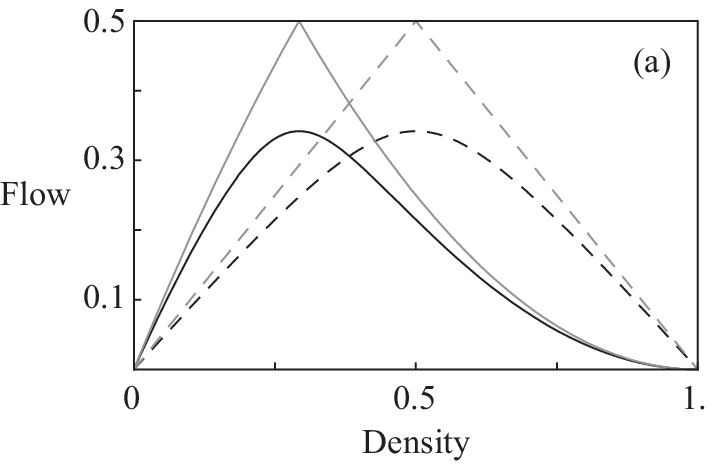}\\
\hspace{6pt}\includegraphics[scale=1.]{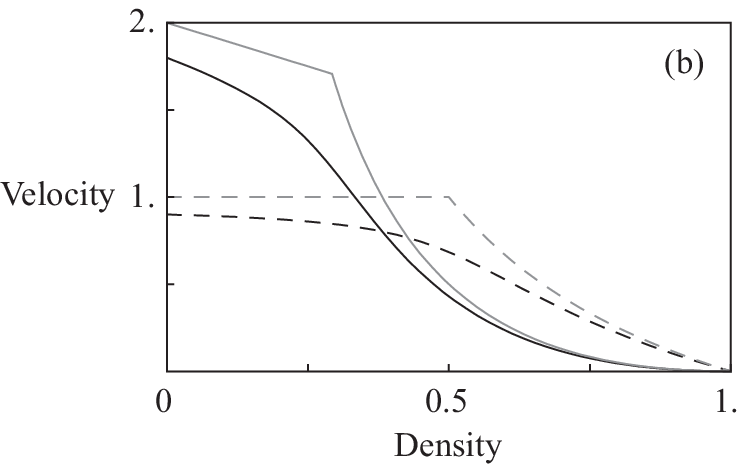}
\caption{
%The flow-density diagrams of the ASEP with the parallel updating and hop probability $p=0.9$, and those of the rule-184 CA ($p=1$); calibrated (black: ASEP, gray: Rule-184 CA), non-calibrated (dashed black: ASEP, gray dashed: Rule-184 CA).
The ASEP with the parallel updating and hop probability $p=0.9$, and those of the rule-184 CA ($p=1$): (a)The flow-density diagram and (b)The velocity-density diagram;
% (black: ASEP, gray: Rule-184 CA)calibrated and (dashed black: ASEP, gray dashed: Rule-184 CA)non-calibrated.
(black solid line)ASEP calibrated, (gray solid line)Rule-184 CA calibrated, (black dashed line)ASEP non-calibrated, (gray dashed line)Rule-184 CA non-calibrated.
}
\label{fig2}
%\end{center}
\end{figure}
%%%%%%%%%

%%%%%%%%%
We add another example: the ASEP with the parallel updating which includes the rule-184 CA as a special ($p=1$).
The flow-density diagram for the parallel update case, which is described as~\cite{Schadschneider}
\begin{equation}
Q_{\mbox{\tiny parallel}}(\rho_{\mbox{\tiny CA}})=\frac12\bigl[1-\sqrt{1-4p\rho_{\mbox{\tiny CA}}(1-\rho_{\mbox{\tiny CA}})}\bigr]
\end{equation}
is calibrated, using eq. (\ref{calib}), to be
\begin{equation}
Q_{\mbox{\tiny parallel}}(\rho_{\mbox{\tiny RW}})=\frac{1-\sqrt{1-4p\rho_{\mbox{\tiny RW}}(2-\rho_{\mbox{\tiny RW}})(1-\rho_{\mbox{\tiny RW}})^2}}{2}.\label{ASEPpc}
\end{equation}
In Fig. 2(a), we show the graph of eq. (\ref{ASEPpc}).
The maximum-flow density calibrated is equal to $1-1/\sqrt{2}$, and is independent of $p$.
Moreover, it is identical to that for the random sequential update case.
In the special case of the rule-184 CA, we have
\begin{equation}
\begin{aligned}
Q_{\mbox{\tiny parallel}}(\rho_{\mbox{\tiny CA}})=&\min\{\rho_{\mbox{\tiny CA}},\,1-\rho_{\mbox{\tiny CA}}\},\\
Q_{\mbox{\tiny parallel}}(\rho_{\mbox{\tiny RW}})=&\min\{\rho_{\mbox{\tiny RW}}(2-\rho_{\mbox{\tiny RW}}),\,(1-\rho_{\mbox{\tiny RW}})^2\}.
\end{aligned}
\label{Qp}
\end{equation}
In this case ($p=1$ and the parallel updating), we see a first-order phase transition at the maximum-flow density, which distinguishes the model from the other ones.
The phase transition implies that a traffic jam never occurs until the density exceeds the maximum-flow density.
(By contrast, we find a traffic jam occurring at rather lower densities in the other cases.)

As well as in the previous case, the average velocity is obtained from the flow via $v=Q/\rho$.
In particular, from eq. (\ref{Qp}), those for the rule-184 CA are given by
\begin{equation}
\begin{aligned}
v_{\mbox{\tiny parallel}}(\rho_{\mbox{\tiny CA}})=&\min\{1,\,-1+1/\rho_{\mbox{\tiny CA}}\},\\
v_{\mbox{\tiny parallel}}(\rho_{\mbox{\tiny RW}})=&\min\{2-\rho_{\mbox{\tiny RW}},\,(1-\rho_{\mbox{\tiny RW}})^2/\rho_{\mbox{\tiny RW}}\}.
\end{aligned}
\label{ASEPvp}
\end{equation}
%Note that the average velocity is no longer equivalent to the mean hop probability of the ASEP.
Figure 2(b) shows the graphs of eq. (\ref{ASEPvp}), which presents how the proposed calibration works.
% the dynamics mimicked by the simple particle-hopping model with a constant hop probability and moreover how the proposed calibration works.
Once the cell size is fixed, dynamics of the {\it real-world} vehicles is transformed into a coarse-grained motion in the CA model: Particles therein hop at most one site, with a constant probability and the exclusive interaction.
The calibration then helps one restore the original dynamics.
Compare Figs. 1 and 2 with the simulated/observed diagrams given in refs. 7, 8, 21, 23, and 24 to see the effect of the present calibration.

%%%%%%%%
%In this letter, we introduced density dependence of the cell size in CA models for traffic flow and showed that the proposed calibration is valid in that it reproduces a realistic flow-density diagram for some basic models.
%We need however to modify the models in order to reproduce detailed structures of traffic flow, such as a metastable branch appearing in the observed flow-density diagram.
%Nevertheless, as far as traffic flow is regarded as a fluid, i.e., on a macroscopic level, these simple CA models share a lot with a real-world traffic flow.
%We conclude, more precisely, that with a calibration of the density, the exclusion rule in CA models shows, beyond the hole-particle symmetry, an asymmetric flow-density relation peculiar and fundamental to traffic flow.
%
%One of our motivations in the present work is to reduce the cost of large-scale traffic simulations such as city traffic and road networks, using some simple model which reproduces a realistic traffic flow.
%Our central idea is that in CA models the cell size, which is usually set to the vehicle size, may also depend on other parameters, e.g., the number of vehicles conserved in each simulation.
%It is not difficult to apply this idea for general CA models such as two-dimensional CA models for pedestrian flow and computation techniques (e.g., the Monte-Carlo method).
In this letter, we proposed a calibration of the particle density in CA models for traffic flow.
%One of our motivations in the present work is to reduce the cost of large-scale traffic simulations such as city traffic~\cite{Gershenson}, using some simple model which reproduces a realistic traffic flow.
One of our motivations in the present work is to reduce the cost of large-scale traffic simulations such as city traffic, using some simple model which reproduces a realistic traffic flow.
Our central idea is that the cell size, which is usually set to the vehicle size, may also depend on other parameters, e.g., the number of vehicles conserved in each simulation.
It is not difficult to apply this idea for general CA models such as two-dimensional CA models for pedestrian flow and computation techniques (e.g., the Monte-Carlo method).
We need however to modify the models in order to reproduce detailed structures of traffic flow, such as a metastable branch appearing in the observed flow-density diagram.
Nevertheless, as far as traffic flow is regarded as a fluid, i.e., on a macroscopic level, these simple CA models share a lot with a real-world traffic flow.
We conclude, more precisely, that with a calibration of the particle density, the exclusion rule in CA models shows, beyond the hole-particle symmetry, an asymmetric flow-density relation peculiar and fundamental to traffic flow.
Finally, it is fair to comment that the calibration method proposed works reliably only for homogeneous systems such as a periodic lattice.
However, we believe that we modify the present method so that one can apply it to the ASEP with an open boundary condition.
It will be reported in our subsequent publications.
%%%%%%%%%

The author is supported by Global COE Program, ``The research and training center for new development in mathematics,'' at Graduate School of Mathematical Sciences, The University of Tokyo.

%%%%%%%%%%%%%%%%%%%%%%%%%%%%%%%%%%%%%%%%%%%%%%%%%%%

\end{document}